\documentclass{article}
\usepackage[utf8]{inputenc}
\usepackage{amsmath}
\usepackage{amssymb}
\usepackage{graphicx}
\usepackage{longtable}
\usepackage{url}
\usepackage{hyperref}
\usepackage{fancyhdr}
\usepackage[margin=1in]{geometry}
\usepackage{caption}
\begin{document}

\title{\huge Are They Willing to Participate? A Review on Behavioral Economics Approach to Voters Turnout}
\author{Mostafa Raeisi Sarkandiz\\
\small Department of Economics, Business and Statistics, University of Palermo, Italy\\
\small mostafa.raeisisarkandiz@unipa.it}
\date{}
\maketitle

\begin{abstract}
\noindent This article investigates the fundamental factors influencing the rate and manner of Electoral participation with an economic model-based approach. In this study, the structural parameters affecting people's decision making are divided into two categories. The first category includes general topics such as economic and livelihood status, cultural factors and, also, psychological variables. In this section, given that voters are analyzed within the context of consumer behavior theory, inflation and unemployment are considered as the most important economic factors. The second group of factors focuses more on the type of voting, with emphasis on government performance. Since the incumbent government and its supportive voters are in a game with two Nash equilibrium, and also because the voters in most cases are retrospect, the government seeks to keep its position by a deliberate change in economic factors, especially inflation and unemployment rates. Finally, to better understand the issue, a hypothetical example is presented and analyzed in a developing country in the form of a state-owned populist employment plan.
\end{abstract}

\vspace{0.1in}

\bigskip


\section{Introduction}

The election is one of the vital elements of democratic governance, and the most crucial component is the degree of citizen participation in the elections because there is such a belief that high turnout reflects the health of the democratic system (Robbins, 2010). Although a high level of turnout reflects the health of the democratic system, the decline in participation in the developed countries seems to be a threat to their democracy, and there is fear that elections as one of the necessary foundations of the democratic system lose its place especially in the public mind. As a result, governments will lose their popular support, and with the lack of cooperation and participation of citizens, they will find themselves virtually in crisis both domestically or internationally. 
\bigskip

\noindent The critical question is why in some societies, the political participation of citizens is high, but in others, it is not. More precisely, what is the difference in these societies that caused this difference in levels of citizen participation?
\bigskip

\noindent About 150 years ago, Scottish historian Alexander Tytler said about democracy, ``The majority always votes for the candidate who promises more profits than the public treasury'' (Tytler, 1840). This perception of how people behave in elections reflects the vital role which economic factors play in the type and quality of people's votes. After this point of view, many researchers called the person who follows the decision method mentioned aboveas ``Economic Voters''. However, this classification may not be particularly accurate because individuals usually consider several different factors in their decisions that would include economic and non-economic factors. The crucial debate is how economic factors affect citizens' voting and decision making. The economic voters' thinking is based on the assumption that when the economic situation is better, the government will have more success in the election (Reidy et al., 2017).
\bigskip

\noindent In today's world, most old democracies have faced the dilemma of diminishing citizen participation in elections. In Switzerland, for example, citizen participation has declined sharply since 1970 onwards, and now the average participation rateis around 40\% (Lutz, 2007). Numerous studies have shown that Canadian youth turnout who born in the 1970s are about 20 percent less likely to vote than their peers who born in 1945-59 (Blais et al., 2004). Similar studies also have shown the same problem in the United States. The following is a summary of the turnout rates in the U.S. presidential elections:
\bigskip

\begin{table}[h]
\centering
\caption{Turnout Rate in the U.S. Presidential Elections}
\begin{tabular}{|c|c|c|}
\hline
\textbf{Year} & Turnout Rate (\%) & Current/Previous (\%) \\
\hline
2016 & 68.29 & +5.97 \\
2012 & 64.44 & +0.12 \\
2008 & 64.36 & - 6.38 \\
2004 & 68.75 & +7.82 \\
2000 & 63.76 & -3.35 \\
1996 & 65.97 & -15.44 \\
1992 & 78.02 & +7.64 \\
1988 & 72.48 & - 2.88 \\
1984 & 74.63 & -2.48 \\
1980 & 76.53 & N.A. \\
\hline
\end{tabular}
\label{tab:turnout_us_pres}
\end{table}
\begin{center}
\small Source: International IDEA Institute
\end{center}
\bigskip

\noindent The table above shows the difference between the lowest and highest turnout rates in 10 presidential elections, was 14.26\%, in which the lowest turnout belongs to 2000 and the highest one was in 1992. To determine why the participation rate in elections fluctuates, the understanding of how individuals make their decisions is essential. Therefore, in this study, we attempt to answer this fundamental question. 
\bigskip 

\noindent The structure of this paper is as follows. In section 2, some of the most important electoral studies, which the core focus of them are on the socio-economic concepts, are reviewed. In section 3, the primary parameters affecting voter shave been discussed. Also, in section 4, the interaction between eligible voters and the incumbent government is analyzed. Ultimately, in section 5, the conclusion is presented.

\section{Literature Perspective}
This section outlines a few studies on how voters make decisions, focusing on economic issues. The reason for the low number of studies presented is because a large proportion of relevant and similar studies have been cited in other sections of the paper. It should be noted that this section is intended solely to provide the reader with a better understanding of the literature of this field, and further studies in this area are presented in other sections.
\bigskip

\noindent Nezi (2012), by analysis several elections in Greece, concluded voters tend to punish the government in times of weak economic performance and vote for rival groups in the elections. Governments should also expect only the support from the side of their core voters at a time of deep economic crisis (The core voters are people who vote for non-economic reasons and psychologically or directly depend on the government). Killian (2008), in his study, reached the fact when people realize that their financial circumstances are lagging behind the national trend and the overall average of the community, their willingness to participate in the elections increases. This exciting conclusion suggests that the overall impact of adverse economic conditions on individuals' decisions is far greater than favorable economic conditions. 
\bigskip

\noindent Burden \& Wichowsky (2014) also pointed out the asymmetry of people's behavior in the deal with economic conditions and stated that harsh economic conditions would increase election participation (for further discussion see: Kahneman \& Tversky, 1979). Radcliff (1992), by analysis of the electoral process in several industrialized countries, found that adverse economic conditions could decline public participation, while the reverse is true for developing countries. Pacek et al. (2009) achieved similar results for developing countries of Eastern European as well.

\section{Voter decision making: General perspective}
In this section, the most important parameters which shape the voter's beliefs are discussed. It should be mentioned that the socio-economic parameters which are understandable by voters in their routing life just investigated in this session.

\subsection{Voter as a Consumer}
The decision to vote generally does not seem to differ from other decisions that people make during their social lives (Lau \& Redlawsk, 2006). Since every individual willing to maximize its utility, Anthony Downs discussed the behavior of individuals in a general election by defining the voters as rational consumers; in his view, being rational means that the individual seeks to satisfy his needs without concealing his interests. Moreover, the individual will stop his performance when he discovers a mistake in his action (Downs, 1957). Based on this definition, if the costs of participating in the elections outweigh its benefits, one would be reluctant to participate. So, for a potential voter, the utility of voting is calculated as follows:

\begin{equation} \label{eq:delta_u}
\Delta U=PB-C
\end{equation}
\noindent Which P represents the probability that a vote can be decisive, B represents the gain of the individual candidate's victory, and C represents the net cost incurred by the individual in the voting process. In practice, this survey leaves us with a deadlock, because in a general election, the number of people who eligible to participate will be substantial, so P is close to zero. Therefore it always seems $\Delta U$ is negative, and virtually no people are willing to vote. 
\bigskip

\noindent To overcome this impasse, Riker \& Ordeshook (1968), in their study, tried to resolve this problem by adding the parameter D as a civic duty. Furthermore, Uhlaner (1989), besides the civic duty, added the willing of supporting charismatic leaders or politicians in D. Also, Edlin et al., (2007) expanded the theory by separating voters into selfish and social. The process of this analysis is as follows:
\bigskip

\noindent By dividing Factor B into the personal income $B_{\text{self}}$, and the social income $B_{\text{soc}}$, and N be the effective population size, then the term B is redefined as follows:

\begin{equation} \label{eq:b_redefined}
B = B_{\text{soc}} + \alpha N B_{\text{self}}
\end{equation}

\noindent Thus, $\alpha$ as a moderating factor indicates that the benefits to other people is less important than the gain to the individual and, therefore, for the majority of individuals $\alpha<1$. Now, if $\alpha=0$ we will be a selfish voter and if $\alpha>0$ then we will be a social voter. It should be noted that $\alpha$ never will be negative. Obviously, the likelihood of a vote to being decisive is inversely related to the number of voters. So we redefine P as follows:

\begin{equation} \label{eq:p_redefined}
P=\frac{K}{n}
\end{equation}

\noindent Which K represents the level of competitiveness of an election. In this definition, $K=10$ is an acceptable value for an election where the parties are close in a competition, and the lower the level of competition will accompany by the lower value of K. By substituting (\ref{eq:p_redefined}) in (\ref{eq:delta_u}), we will have:

\[
\Delta U=\frac{K}{n}B-C
\]

\noindent By defining $b=\frac{B}{N}$ as expected profit per person:

\begin{equation} \label{eq:b_expected_profit}
b=\alpha B_{\text{soc}}+\frac{1}{N}B_{\text{self}}
\end{equation}

\noindent We will also define $n_{\text{elig}}$ as the number of eligible voters and T as voter turnout ratio, thus $n=T n_{\text{elig}}$:

\begin{equation} \label{eq:delta_u_turnout}
\Delta U=K \frac{N}{n}b-C=K \frac{N}{T n_{\text{elig}}}b-C
\end{equation}

\noindent We know that if N is large enough, then $b\approx\alpha B_{\text{soc}}$, so b must be positive. We know from equation (\ref{eq:delta_u}) that it is rational for a person to participate in elections if $\Delta U>0$ so we have:

\[
\Delta U>0 \quad \text{if}
\]
\begin{equation} \label{eq:rational_condition}
C/b < K \frac{N}{T n_{\text{elig}}}
\end{equation}

\noindent Given a statistical distribution for $c/_{b}$ among the population eligible to vote, we can examine the difference in the turnout in various elections. For minor elections b will decrease and so $c/_{b}$ will increase. As a result, fewer people will be willing to vote. However, because the rate of participation has reduced, this is important for some rational people (marginal people) to enter the polls, so it expects that there will be a balance of participation $(T_{\text{equilib}})$.
\bigskip

\noindent In the views mentioned above, voting only studies within the framework of microeconomics, and macroeconomic factors have not explicitly discussed. Besides, the impact of government activities on citizen participation has not accounted for, as well. The cost factor here is severely limited and covers some of the costs, including the cost of being in the voting queue, the cost of gathering information to select an ideal candidate, etc. While these costs are significant, however, it cannot explain how the increase and decrease in the inflation rate will affect turnout rates.
\bigskip

\noindent We start with the fact that voting is just like other people's choices, and so the principle of maximizing utility function will be in place. The issue of maximizing consumer desirability is as follows:

\begin{equation} \label{eq:max_utility}
\max_{X\in R_{+}^{n}}U(X) \quad \text{s.t} \quad P\cdot X\le Y
\end{equation}

\noindent Where $X=(x_{1},...,x_{n})$ is a vector containing different quantities of all the commodities that one can chose in the bundle of its selection. Also $P_{i}$ represents the price of commodity iand $P=(p_{1},...,p_{n})$ and Y represents income. Thus $X=\Sigma_{i=1}^{n}p_{i}x_{i}$ (Jehle \& Reny, 2011). Therefore, the factors that influence a person's desirability are equal to the factors that include his or her limitation. Since $P\cdot X\le Y$ then these factors will be the price of the commodities and the income level of the individual.
\bigskip

\noindent Let us look at prices. What factors make prices fluctuate? Obviously, inflation is a benchmark for changing prices, and it is also an index that the householders experience directly. Now, if inflation is positive, it means that prices have gone up, so people with a fixed budget will be less able to buy the goods they want, and vice versa. As a result, positive inflation reduces people's utility (Raeisi Sarkandiz \& Bahlouli, 2019). Now let us look at the revenue factor. In general, having a job creates income for people, so the threat of employment can lead to income loss. It is also clear that the lack of employment, will drastically reduce utility, and as a consequence of what society perceives directly, unemployment decreases their well-being.
\bigskip

\noindent In summary, it can be stated that the two main factors that people are facing tangible and affect their desirability directly and profoundly are inflation and unemployment rates. In this regards, Lewis-Beck \& Paldam (2000) on the impact of the economic situation on elections by review numerous studies deduced as follows:
\begin{enumerate}
    \item Economic changes account for about one-third of the changes in voting and election results.
    \item Voters react more to past events than to future expectations. In other words, they look to the past rather than the future.
    \item Inflation and unemployment are the most significant macroeconomic factors affecting the election.
    \item Voters have a short-term horizon.
\end{enumerate}

\subsection{Cultural parameters}
One of the issues that can be addressed in the electoral field is the race or ethnicity of the candidates; for example, the presence of a black candidate, especially in the U.S. elections. Lublin \& Tate (1995) showed turnout increases when a black candidate has placed on the candidate list. Also, Washington (2006) stated there is plausible evidence that black candidates in the United States are increasing election turnout. 
\bigskip

\noindent In some cases, this increase is because black voters have come out in support of black candidates, and other whites have come out in support of other candidates especially the pure white ones. A racial or linguistic ethnicity can increase the participation among the community from which a specific candidate is nominated. The extent to which black persons participated in U.S. elections under President Barack Obama, the below table may support this.

\begin{table}[h]
\centering
\caption{\small Blacks' Turnout in the U.S. Presidential Elections}
\begin{tabular}{|c|c|c|c|c|}
\hline
Gender/Year & \multicolumn{2}{c|}{Eligible Persons (\%)} & \multicolumn{2}{c|}{People Who Report Voting (million)} \\
\cline{2-5}
& Female & Male & Female & Male \\
\hline
2012 & 70.1 & 61.4 & 10.4 & 7.4 \\
2008 & 68.1 & 60.5 & 9.4 & 6.7 \\
2004 & 63.4 & 55.8 & 8.3 & 5.7 \\
1996 & 56.1 & 49.1 & 6.7 & 4.7 \\
1992 & 59.2 & 53.9 & 6.6 & 4.8 \\
1988 & 55.9 & 50.5 & 5.9 & 4.2 \\
1984 & 60.7 & 54.1 & 6.1 & 4.2 \\
\hline
\end{tabular}
\label{tab:blacks_turnout}
\end{table}
\begin{center}
\small Source: CAWP (2015).
\end{center}
\bigskip

\noindent In addition to race or color, the turnout in the local elections is undoubtedly lower than the national ones, and so the turnout for the congressional elections is lower than the presidential.

\begin{table}[h]
\centering
\caption{\small Voter Turnout in The U.S. Elections}
\begin{tabular}{|c|c|c|c|}
\hline
\multicolumn{2}{|c|}{Presidential} & \multicolumn{2}{c|}{Congressional} \\
\hline
Year & Participation Rate & Year & Participation Rate \\
\hline
2016 & 68.29 & 2014 & 42.50 \\
2012 & 64.44 & 2010 & 48.59 \\
2008 & 64.36 & 2006 & 47.52 \\
2004 & 68.75 & 2002 & 45.31 \\
2000 & 63.76 & 1998 & 51.55 \\
1996 & 65.97 & 1994 & 57.64 \\
1992 & 78.02 & 1990 & 56.03 \\
1988 & 72.48 & 1986 & 54.89 \\
1984 & 74.63 & 1982 & 61.10 \\
1980 & 76.53 & 1978 & 57.04 \\
\hline
\multicolumn{2}{|c|}{Average: 69.72} & \multicolumn{2}{c|}{Average: 52.22} \\
\hline
\end{tabular}
\label{tab:turnout_congress_pres}
\end{table}
\begin{center}
\small Source: International IDEA Institute
\end{center}

\subsection{Psychological factors}
The Bandwagon effect is a situation where a person cut off or raise his demand based on the demand of other people in the market (Leibenstein, 1950). In other words, if the demand of the people in the market increases, he will increase his demand and vice versa (Raeisi Sarkandiz, 2020). The interpretation of this rule in the electoral field is that voters are more vote for a candidate who is more likely to succeed in the election (Kiss \& Simonovits, 2014). The effect of this phenomenon on the voting process has been generally demonstrated (Zech, 1975). However, what is the mechanism of this effect, and how does it affect the outcome of the vote? 
\bigskip

\noindent Many recent empirical studies have found acceptable evidence that this effect has increased participation rates in support of the significant candidate (see, for example, Klov \& Winter, 2007; Grosser \& Schram, 2010; Agranov et al., 2017). During the 2000 U.S. presidential election, the national media mistakenly declared that voting was over in Florida and that the Democratic Party had won in that state, while polling stations in some western areas were still receiving votes. In these areas, Lott (2005) recorded a sharp decline in the turnout of the Republican supporters over the Democrats following that misinformation (Grillo, 2017).
\bigskip

\noindent Generally, a president who is chosen for a period may win the next run. This outcome is likely due to the Bandwagon effect and thus makes the current government's opponents less likely to vote and lead to lower turnout. Of course, this cannot be stated explicitly and requires further study and research because, in some countries and some periods, this is not a fixed process. In this regard, the tendency of the presidential election in the United States does not show the same trend, while we see this effect clearly in the presidential elections of Iran (except in 2009, when the massive social changes increased the turnout).

\begin{table}[h]
\centering
\caption{Voter Turnout in the U.S. Elections by President}
\begin{tabular}{|c|c|c|c|}
\hline
Year & Participation Rate & President & Turnout Change \\
\hline
2012 & 64.44 & B. Obama & Increase \\
2008 & 64.36 & & \\
\hline
2004 & 68.75 & G. W. Bush & Increase \\
2000 & 63.76 & & \\
\hline
1996 & 65.97 & B. Clinton & Decrease \\
1992 & 78.02 & & \\
\hline
1984 & 74.63 & R. Reagan & Decrease \\
1980 & 76.53 & & \\
\hline
\end{tabular}
\label{tab:us_pres_by_pres}
\end{table}
\begin{center}
\small Source: International IDEA Institute
\end{center}
\bigskip

\begin{table}[h]
\centering
\caption{Voter Turnout in Iran Elections by President}
\begin{tabular}{|c|c|c|c|}
\hline
Year & Participation Rate & President & Turnout Change \\
\hline
1981 & 74.26 & S. A. Khamenei & Decrease \\
1985 & 54.78 & & \\
\hline
1989 & 54.59 & A. Rafsanjani & Decrease \\
1993 & 50.66 & & \\
\hline
1997 & 79.92 & S. M. Khatami & Decrease \\
2001 & 66.77 & & \\
\hline
2005 & 62.84 & M. Ahmadi Nejad & Increase \\
2009 & 84.83 & & \\
\hline
\end{tabular}
\label{tab:iran_pres_by_pres}
\end{table}
\begin{center}
\small Source: Iran Election Headquarters
\end{center}
\bigskip

\noindent It is clear that the number of votes cast in the first round is usually more than the number of votes cast in the second round, except for the period in which massive political, economic, or national changes happened (for instance, September 11 terrorist attacks in the U.S.).

\section{Government vs. Voters}
Each citizen in the face of the incumbent government can show three behaviors includes: agree, disagree, or indifference. An indifferent person will not be willing to participate in the elections. Nevertheless, if the economic performance of the government is positive, the person who was the opposite of the government will lose its willingness to participate in the elections. If his/her economic condition gets worse than in the past, he/she will participate in the elections in opposition to the government. It is useful here to consider a well-known concept, the Bandwagon Effect because it can have a profound effect on the behavior of those who vote.
\bigskip

\noindent Now, suppose the economic situation has improved, and it is time for the second round of the election. The incumbent government's behavior will subject to numerous cases. The government's priority is re-electing, so the government and its supporters are practically in a game. In any game, each side seeks to make the best move. On the other hand, we know that each game has at least one equilibrium known as Nash equilibrium in the game theory. The Nash equilibrium is the condition in which each player chooses a strategy that, concerning the other strategies, gives him/her the most revenue. 
\bigskip
\noindent Of course, in some games, there is more than one equilibrium. Since neither side knows the strategy of the other, so the game will be simultaneous (Besanko \& Braeutigam, 2014). The following is the game of the government and its supporters for the next round of elections if the economic conditions got improved:
\bigskip

\noindent Given supportive voters face a participation cost, they prefer not to turn out even when the incumbent is expected to win. The government, whose priority is re-election, strictly prefers a victory outcome. Representing payoffs as (supportive voter, government), the game

\begin{center}
\begin{tabular}{c|cc}
 & Victory & Failure \\
\hline
Participation & (0,2) & (-1,0) \\
Non-Participation & (1,2) & (0,0)
\end{tabular}
\end{center}

\noindent shows that non-participation strictly dominates participation for the supportive voter, while victory strictly dominates failure for the government. The unique Nash equilibrium is therefore (Victory, Non-Participation): the incumbent wins while supportive voters abstain because the cost of participation outweighs its benefit.
\bigskip

\noindent Therefore, the government makes its efforts to motivate supporters to participate in the elections, while supportive voters prefer that the government re-elect at no cost. This interplay explains much of the turnout rates in a general election. The more successful the government is in this game, the higher the participation rate, and vice versa. Therefore, first, the government will try foremost to make re-vote all who voted for him in the previous round, and second, attempt to reduce the votes of its rivals. However, the government's performance makes it even more likely that some people who did not want to participate or were indifferent to the government, motivated to participate in the election.
\bigskip

\noindent We now review the tactics used by the government to increase the turnout. The first step of the government will be to reduce the direct cost of participating in the elections for all eligible voters.

\subsection{Direct Costs Reduction}
As mentioned earlier, voting is in the bundle of people's choices and situated among the consumer goods. Direct election costs include the cost of getting off of home or work to the polls, the cost of the stand in the voting queue (time cost), as well as the cost of gathering information to select the person from the visual, audio, and print media or being present at the site of advertising campaigns. Although decreasing these costs can lead to more people's willingness to participate in the elections, but eliminating all the direct costs cannot lead to full turnout. However, it is noteworthy that the establishment of various methods, including postal voting, electronic voting, and the provision of new services to people with disabilities, generally have a positive impact on voter turnout. 
\bigskip

\noindent For instance, Surveys show that the introduction of the postal voting system in Switzerland has increased citizen participation by about four percent (Luechinger et al., 2007). Similar effects have also been observed in other countries (Gronke et al., 2008). It should be noted that the impact of internet voting on participation cannot be entirely ascertained because, in some areas, this method made increased, in some others made decreased, and for the rest of them, that was insignificant. Therefore, this will require further investigation (Germann \& Serdult, 2017). Miller \& Powell (2015) found in their survey of the relationship between disability and election participation, people with disabilities were less likely to vote than ordinary people and also found that people with disabilities were more inclined to vote by the mail. 
\bigskip

\noindent Therefore, establishing the appropriate conditions for them will reduce the direct costs of voting and will increase the likelihood of participating. It is clear, the efforts to reduce the cost of the participants can have a positive effect, but it cannot see as an essential factor in assessing the extent to which people participate in the political process. There are several other factors that each of them can somehow influence the turnout, although we reiterate that these factors alone cannot explain the rate of participation in an election, but together they will be able to explain the rate.

\subsection{Hyperbolic Memory Discount Effect}
For the first time, Nordhaus (1975) stated the government by deliberate changes in macroeconomic indices, would encourage the citizens to vote again. In days leading the election-day, the government will pursue a policy that reduces the inflation rate, while a high unemployment rate would be a collateral outcome. Then, the unemployment rate begins to fall back to their optimum levels during the period and will shortly continue after the election, following the implementation of inflationary policies and the suspension of those policies. 
\bigskip

\noindent Once the government is re-elected, the trend will repeat, and unemployment will rise again as inflation drops. In this process, the government uses fiscal/budgetary policies as monetary policies. The Philips curve could illustrate the short-term trade-off between inflation and unemployment. Therefore, the behavior of the government merely moves the Phillips curve in the short run. The reason the government adopts this behavior is that it faces retrospect voters who are experienced macroeconomic indicators such as high unemployment and low inflation. Voters in this context are exponentially discounting their memory of problems arising from inflation and unemployment (Findley, 2015). For example, the government target inflation over unemployment rates and vice versa.

\subsection{Size of the Government}
The size of the government and the number of people directly employed by the government has a significant relationship with the turnout. The government employees are more likely to vote in favor of the government because it has a direct relationship with their economic benefits. Also, when we consider these people as a voting bulk, we find out that their influence on the election is more prominent than their number (Bennett \& Orzechowski, 1983). To explain this phenomenon, Borcherding et al., (1977) defined the "voting power index" as follows:

\begin{equation} \label{eq:voting_power_index}
P = \frac{1}{1 + \frac{V_{NB}}{V_B} (\frac{1}{G} - 1)}
\end{equation}

\noindent Which $V_B$ represents the participation rate of bureaucrats in the elections, $V_{NB}$ represents the turnout rate of the non-bureaucrats and $G$ is the percentage of the total number of government employees in the labor market. Taking $V_B = 0.9$ and $V_{NB} = 0.5$, Bush \& Denzau (1977) showed that if $G = 5\%$, their overall effect is 8\%. If $G = 10\%$ then their effect is 16\% and if $G = 40\%$ then their effect will be 51\%. Therefore, it can be seen how a minority can become the majority. 
\bigskip

\noindent In other words, bureaucrats have more power than they can count because they are always more likely to participate in elections than non-bureaucrats ones. So, they constitute a higher proportion of voters (Corey \& Garand, 2002). Therefore, the higher the number of government employees, the higher the turnout rate in favor of the incumbent government, which explains why governments are always reluctant to downsize. Looking at the process of development and economic growth in developed countries, we find that during the period, the size of the government has decreased and consequently the number of government employees has decreased over time, so this block and the voting group have declined sharply, so $P$ has gradually dropped. By using the theory of voting power index, one can deduce that one of the reasons for the decline in turnout rate in developed countries is the decrease in the volume and number of government employees.
\bigskip

\noindent We have seen that increasing the number of government employees will lead to an increase in participation and it can be account as a chief factor in the government's re-election. So how can the government increase its staff? The first point is that the increase in staff should occur in the year leading up to the election. Second, since the plan cannot finance directly (the parliament usually will not allow such populist projects), the government provides resources by creating unpredictable and in some cases, illegal deficits. However, these resources will eventually have to be sorted out in some way, because if they are not settled, it will be possible to pursue justice through competent authorities. The government plan to overcome this budget deficit includes two feasible options which describe as follows.

\subsubsection{Post-election Job Cuts}
Suppose the government recruit new workers and fund the job position through a budget deficit, which equals to $X$. The employment contract is generally one-year and it will expire after the election. The new employees tend to extend their contracts for the next year, but the government's goal is to attract votes, with no willingness to pay after re-election. As a result, the contracts will not be renewed in the first place. 
\bigskip

\noindent On the other hand, the government needs to apply a cost reduction policy to offset the financial cost of $X$. Therefore, the government has to cut off a number of its previous jobs, and consequently, the employee reduction will offset the cost of $X$. Besides, due to the effect of multiplication or multiplier, the budget deficit will exceed its original value, so assuming $Y$ as the surplus. Hence, some jobs need to be eliminated again to offset the financial cost of $Y$; therefore we will have several jobs eliminated, and this reduction is more than the absorbed number. The behavior of the government in this way illustrated as follows:

\subsubsection{Preservation of the Created Jobs}
In this scenario, if the government willing to maintain the created jobs even after winning the election, one of the solutions would be to use the Ponzi Scheme. According to this method, the government issues bonds for the expenses incurred. As the government does not have the financial resources to pay for these financial securities, it will also issue new bonds to repay their interest and will repeat the process several times (Bartolini \& Cottarelli, 1994). 
\bigskip

\noindent This process cannot go on like this. Because the issuance of multiple securities leads to an increase in interest rates and consequently will lead to a reduction in investment that can cause a financial crisis. Under these circumstances, the job reduction will occur in the non government section. In developing countries, the government can offset the mentioned deficit by adopting policies such as currency exchange or selling its assets. However, we know that financing from currency exchange will lead to market volatility, and the sale of assets also results in a reduction in government employees. 
\bigskip

\noindent Therefore, we can conclude that after the implementation of this plan, many job positions will be eliminated, either by the government or by the non-governmental sector. Of course, by supposing the hyperbolic memory discounting effect mentioned earlier, this policy could work for the government and provide part of its votes for re-election. Finally, it is worth noting that this policy is not possible in countries where parliamentary or judiciary oversight is continuous and pragmatic. So, the government is required to be transparent. As a result, this plan maybe implemented in developing or underdeveloped countries.

\section{Concluding remark}
In this article, the role of economic factors that affectthe citizen's decision in a local or national election has been studied. Given that the electoral process needs to be analyzed in a socio-economic context, therefore, it is best to use a behavioral economics approach, since this approach has increased the flexibility by incorporating psychological parameters into economic models. The variables analyzed in this study are divided into two categories. 
\bigskip

\noindent The first category was related to the factors that generally affect voters (regardless of their voting type). Taking a theory-based approach to consumer behavior, we found that the most significant factors in voters' decision-making are inflation and unemployment rates. The second section was focused on the impact of government actions on citizens' electoral behavior. It has been found that we generally face voters who want to minimize their costs and tend to punish the government in severe economic conditions. 
\bigskip

\noindent In the case of historical processes, they also exhibit retrospective behavior. In the meantime, the government is seeking re election, as well as pro-government voters (if the country's economic situation is favorable) are seeking re-election of the government at the lowest individual cost. So the government and its supporters will enter into a game with two Nash equilibrium. 
\bigskip

\noindent Another key factor affecting the citizens'participation rate and also the government's re-election is the number of public employees (bureaucrats). Numerous studies have shown that the effectiveness of the government employees in determining the outcome of an election is higher than their number. This fact may explain why most governments resist structural downsizing and reducing their direct labor forces. 
\bigskip

\noindent Finally, by providing a hypothetical example in a developing country, the issue was further elaborated. The government plan is to hire many employees (nationwide) before the election. However, after the election, and in the event of re election, the government faces two scenarios of removing or extending the contract of newlyhired employees. The analysis of the consequences of the two scenarios showed that if either of these scenarios were implemented, the economic structure of the country and labor market would face extremely devastating results; while the chances of government re-election will increases.

\section*{Declaration}
The author declares no conflict of interest.
\bigskip

\noindent This research did not receive any specific grant from funding agencies in the public, commercial, or not-for-profit sectors.

\section*{References}
\begin{enumerate}
    \item Agranov, M., Goeree, J., Romero, J., \& Yariv, L. (2017). What makes voter turnout:The effect of polls and beliefs. Journal of European Economic Association, 16(3), 825-856. doi. 10.1093/jeea/jvx023
    \item Bartolini, L., \& Cottarelli, C. (1994). Government Ponzi games and the sustainability of public deficits under uncertainty. Ricerche Economiche, 48(1), 1-22. doi. 10.1016/0035-5054(94)90017-5
    \item Bennett, J.T., \& Orzechowski, W.P. (1983). The voting behavior of bureaucrats:Some empirical evidence. Public Choice, 41, 271-283. doi. 10.1007/BF00210361
    \item Besanko, D., \& Braeutigam, R. (2014). Microeconomics (Fifth Edition). John Wiley \& Sons, Inc.
    \item Blais, A., Nadeau, R., Gidengil, E., \& Nevitte, N. (2004). Where does turnout decline come erom?, European Journal of Political Research, 43(2), 221-236. doi. 10.1111/j.1475-6765.2004.00152.x
    \item Borcherding, T.E., Bush, W.C., \& Spann, R.M. (1977). The effects on public spending of the divisibility of public outputs in consumption, bureaucrats: Power, and the size of the tax sharing group. In, Budgets and Bureaucrats: The Source of Government Growth, T.E. Borcherding (Edt), Durham: Duke University Press.
    \item Burden, B.C., \& Wichowsky, A. (2014). Economic discontent as a mobilizer: Unemployment and voter turnout. The Journal of Politics, 76(4), 887-898. doi. 10.1017/S0022381614000437
    \item Bush, W.C., \& Denzau, A.T. (1977). The voting behavior of bureaucrats and public sector growth. In, Budgets and Bureaucrats: The Source of Government Growth, T.E. Borcherding (Edt), Durham: Duke University Press.
    \item CAWP. (2015). Gender Differences in Voter Turnout. Center for American Women and Politics, CAWP.
    \item Corey, E.C., \& Garand, J.C. (2002). Are government employees more likely to vote: An analysis of turnout in the 1996 U.S. national election. Public Choice, 111, 259-283. doi. 10.1023/A:1015290806607
    \item Downs, A. (1957). An Economic Theory of Democracy. New York: Harper and Row.
    \item Edlin, A., Gelman, A., \& Kaplan, N. (2007). Voting as a rational choice. Rationality and Society, 19(3), 293-314. doi. 10.1177/1043463107077384
    \item Findley, T.S. (2015). Hyperbolic memory discounting and the political business cycle. European Journal of Political Economy, 40(B), 345-359. doi. 10.1016/j.ejpoleco.2015.08.002
    \item Germann, M., \& Serdult, U. (2017). Internet voting and turnout: Evidence from Switzerland. Electoral Studies, 47, 1-12. doi. 10.1016/j.electstud.2017.03.001
    \item Grillo, A. (2017). Risk aversion and bandwagon effect in the pivotal voter model. Public Choice, 172(3), 465-482. doi. 10.1007/s11127-017-0457-5
    \item Gronke, P., Galanes-Rosenbaum, E., Miller, P.A., \& Toffey, D. (2008). Convenience voting. Annual Review of Political Science, 11, 437-455. doi. 10.1146/annurev.polisci.11.053006.190912
    \item Grosser, J., \& Schram, A. (2010). Public opinion polls,voter turnout,and welfare:An experimental study. American Journal of Political Science, 54(3), 700-717. doi. 10.1111/j.1540-5907.2010.00455.x
    \item International IDEA Institute, (2020). [Retrieved from].
    \item Iran Election Headquarters, (2021). [Retrieved from].
    \item Jehle, G.A., \& Reny, P.J. (2011). Advanced Microeconomic Theory. London: Pearson Education.
    \item Kahneman, D., \& Tversky, A. (1979). Prospect theory: An analysis of decision under risk. Econometrica, 47(2), 263-292. doi. 10.2307/1914185
    \item Killian , M., Schoen, R., \& Dusso, A. (2008). Keeping up with the Joneses: The interplay of personal and collective evaluations in voter turnout. Political Behavior, 30(3), 323-340. doi. 10.1007/s11109-007-9051-8
    \item Kiss, A., \& Simonovits, G. (2014). Identifying the bandwagon effect in two-round elections. Public Choice, 160, 327-344. doi. 10.1007/s11127-013-0146-y
    \item Klov, E.F., \& Winter, E. (2007). The welfare effects of public opinion polls. International Journal of Game Theory, 193, 379-394. doi. 10.1007/s00182-006-0050-5
    \item Lau, R.R., \& Redlawsk, D.P. (2006). How Voters Decide. Cambridge University Press.
    \item Leibenstein, H. (1950). Bandwagon, snob, and Veblen effects in the theory of consumer's demand. The Quarterly Journal of Economics, 64(2), 183-207. doi. 10.2307/1882692
    \item Lewis-Beck, M.S., \& Paldam, M. (2000). Economic voting: An introduction. Electoral Studies, 19(2-3), 113-121. doi. 10.1016/S0261-3794(99)00042-6
    \item Lott, J.R. (2005). The impact of early media election calls on Republican voting rates in Florida's Western Panhandle counties in 2000. Public Choice, 120, 349-361. doi. 10.1007/s11127-005-7166-1
    \item Lublin, D., \& Tate, K. (1995). Racial group competition in urban election. In P.E. Peterson (Edt), Classifying by Race. Princeton: Princeton University Press.
    \item Luechinger, S., Rosinger, M., \& Alois, S. (2007). The impact of postal voting on participation: Evidence for Switzerland. Swiss Political Science Review, 13(2), 167-202. doi. 10.1002/j.1662-6370.2007.tb00075.x
    \item Lutz, G. (2007). Low turnout in direct democracy. Electoral Studies, 26(3), 624-632. doi. 10.1016/j.electstud.2006.10.008
    \item Miller, P., \& Powell, S. (2015). Overcoming,voting obstacles: The use of convenience voting by voters with disabilities. American Politics Research, 44(1), 1-28. doi. 10.1177/1532673X15586618
    \item Nezi, R. (2012). Economic voting under the economic crisis: Evidence from Greece. Electoral Studies, 31(3), 498-505. doi. 10.1016/j.electstud.2012.02.007
    \item Nordhaus, W.D. (1975). The political business cycle. Review of Economic Studies, 42(2), 169-190. doi. 10.2307/2296528
    \item Pacek, A.C. Pop-Eleches, G., \& Tucker, J.A. (2009). Disenchanted or discerning: Voter turnout in post-communist countries. Journal of Politics, 71(2), 473-491. doi. 10.1017/s0022381609090409
    \item Radcliff, B. (1992). The welfare state, turnout, and the economy: A comparative analysis. American Political Science Review, 86(2), 444-454. doi. 10.2307/1964232
    \item Raeisi Sarkandiz, M., \& Bahlouli, R. (2019). The Stock Market between Classical and Behavioral Hypotheses: An Empirical Investigation of the Warsaw Stock Exchange, Econometric Research in Finance, 4(2), 67-88. doi. 10.33119/ERFIN.2019.4.2.1
    \item Raeisi Sarkandiz, M. (2020). Is this essential for Japan to changes its LNG import policy? Some evidence of the OPEC crude oil price shocks, Economics, Management and Sustainability, 5(1), 29-41, doi. 10.14254/jems.2020.5-1.3
    \item Reidy, T., Suiter, J., \& Breen, M. (2017). Boom and bust: Economic voting in Ireland. Politics, 38(2), 1-17. doi. 10.1177/0263395716680827
    \item Riker, W.H., \& Ordeshook, P.C. (1968). A theory of the calculus of voting. The American Political Science Review, 62(1), 25-42. doi. 10.2307/1953324
    \item Robbins, J.W. (2010). The personal vote and voter turnout. Electoral Studies, 29(4), 661-672. doi. 10.1016/j.electstud.2010.07.001
    \item Tytler, A. (1840). Universal History,From the Creation of the World to the Eighteenth Century. Boston: Hilliard,Gray and Co.
    \item Uhlaner, C.J. (1989). Rational turnout: The neglected role of groups. American Journal of Political Science, 33(2), 390-422. doi. 10.2307/2111153
    \item Washington, E. (2006). How black candidates affect voter turnout. The Quarterly Journal of Economics, 121(3), 973-998. doi. 10.1162/qjec.121.3.973
    \item Zech, C.E. (1975). Leibenstein's effect as applied to voting. Public Choice, 21, 117-122. doi. 10.1007/BF01705954
\end{enumerate}

\end{document}